\documentclass[%
 reprint,
superscriptaddress,
 amsmath,amssymb,
 aps,
 pra,
]{revtex4-1}

\usepackage{graphicx}

\newcommand{\vac}{\varnothing}
\newcommand{\ket}[1]{|#1\rangle}
\newcommand{\bra}[1]{\langle#1|}

\begin{document}

\title{Dressed states of a quantum emitter strongly coupled to a metal nanoparticle}

\author{H. Varguet}
\affiliation{Laboratoire Interdisciplinaire Carnot de Bourgogne, CNRS UMR 6303, Universit\'e de Bourgogne,
BP 47870, 21078 Dijon, France}
\author{B. Rousseaux}
\affiliation{Laboratoire Interdisciplinaire Carnot de Bourgogne, CNRS UMR 6303, Universit\'e de Bourgogne,
BP 47870, 21078 Dijon, France}
\author{D. Dzsotjan}
\affiliation{Laboratoire Interdisciplinaire Carnot de Bourgogne, CNRS UMR 6303, Universit\'e de Bourgogne,
BP 47870, 21078 Dijon, France}
\affiliation{Wigner Research Center for Physics, Hungarian Academy of Sciences, Konkoly-Thege Miklos ut 29-33, H-1121 Budapest, Hungary}
\author{H. R. Jauslin}
\affiliation{Laboratoire Interdisciplinaire Carnot de Bourgogne, CNRS UMR 6303, Universit\'e de Bourgogne,
BP 47870, 21078 Dijon, France}
\author{S. Gu\'erin}
\affiliation{Laboratoire Interdisciplinaire Carnot de Bourgogne, CNRS UMR 6303, Universit\'e de Bourgogne,
BP 47870, 21078 Dijon, France}
\author{G. Colas des Francs}
\email{gerard.colas-des-francs@u-bourgogne.fr}

\affiliation{Laboratoire Interdisciplinaire Carnot de Bourgogne, CNRS UMR 6303, Universit\'e de Bourgogne,
BP 47870, 21078 Dijon, France}

\begin{abstract}
Hybrid molecular-plasmonic nanostructures have demonstrated their potential for surface enhanced spectroscopies, sensing or quantum control at the nanoscale. In this work, we investigate the strong coupling regime and explicitly describe the hybridization between the localized plasmons of a metal nanoparticle and the excited state of a quantum emitter, offering a simple and precise understanding of the energy exchange in full analogy with cavity quantum electrodynamics treatment and dressed atom picture. Both near field emission and far field radiation are discussed, revealing the richness of such optical nanosources.  
\end{abstract}


\maketitle

Optical microcavities can store light for a long time allowing efficient light-matter interaction with important applications in quantum technologies, low threshold laser \cite{Nomura-Arakawa:2008}, supercontinuum laser \cite{Grelu:2016} or indistinguishable single photon source \cite{Laurent-Abram:2005}. It relies on the extremely high quality factor of the cavity mode but at the price of diffraction limited sizes. That is why strong efforts have be done since a decade to transpose cavity quantum electrodynamics (cQED) concepts to nanophotonics and plasmonics \cite{Chang-Sorensen-Hemmer-Lukin:2006,Cuche2010,Tame-Maier:2013,Benson:2014,GCF-Barthes-Girard:2016}. Particular attention has been devoted to the strong coupling regime \cite{Truegler-Hohenester:2008,AbreeGuebrou-Belessa:2012,Delga-GarciaVidal:2014,Zengi-Kall-Shegai:2015} since it offers the possibility of a control dynamics of the light emission, as \emph{e.g.} photon blockade \cite{Smolyaninov-Zayats-Gungor-Davis:2002,Alpeggiani-Gerace:2016} or coherent control \cite{Fleischhauer:2010,Rousseaux-GCF:2016}. 

In this letter, we build an effective Hamiltonian that fully transposes the cQED description to an hybrid plasmon-quantum emitter nanosource. We demonstrate it  can be exactly described in full analogy with cQED representation. Specifically, the coupled plasmon-emitter system behaves like an emitter in a multimodal lossy cavity. We notably determine the structure of the emitter states dressed by the plasmon modes.

\begin{figure}[h!]
\centering
\includegraphics[width=0.35\textwidth]{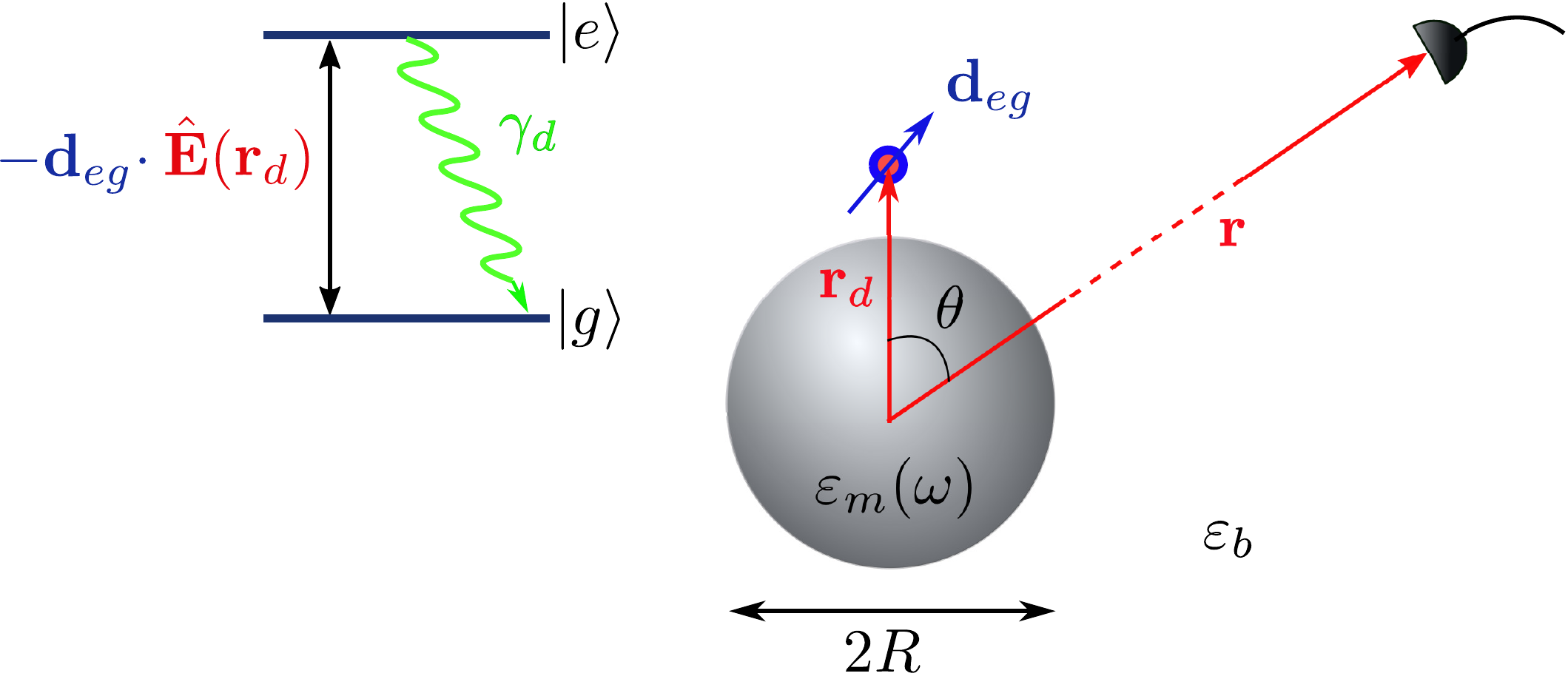}\caption{Scheme of the hybrid system embedded in a background material with permittivity $\epsilon_b=1$.} \label{MNP} 
\end{figure}

We consider the hybrid system displayed in Fig. \ref{MNP}. A two level system (TLS) quantum emitter is located close to a metal nanoparticle (MNP). The optical transition is characterized by the frequency $\omega_{eg}$, the dipole moment $\mathbf{d}_{eg}$ and the operator $\hat{\sigma}_{eg}^{\dagger}=\ket{g}\bra{e}$. For the sake of clarity, we consider a TLS emitter coupled to spherical MNP since the localized surface plasmon (LSP) modes involved in the coupling process are well identified and the hybridization of the emitter and MNP modes will be unambiguously demonstrated. 
%

The Hamiltonian of the coupled system writes
\begin{align}
\hat H=&\sum_{i=e,g}\hbar\omega_{i}\hat{\sigma}_{ii}-i\hbar\frac{\gamma_{d}}{2}\hat{\sigma}_{ee}
+\int d\mathbf{r} \int_0^{+\infty}\!\!\!\!\!\!\! d\omega\ \hbar\omega\  
\hat{\mathbf{f}}^{\dagger}(\mathbf{r},\omega)\cdot\hat{\mathbf{f}}(\mathbf{r},\omega)\nonumber\\
&-\left[\hat{\sigma}_{eg} \int_0^{+\infty}\!\!\!\!\!\!\! d\omega\ \mathbf{d}_{eg}\cdot\hat{\mathbf{E}}(\mathbf{r}_d,\omega)+H.c.\right]\label{hamil}
\end{align}
The first term refers to the TLS energy and we have phenomelogically introduced the decay rate of the excited state $\gamma_d$ in the second term.
The third term describes the total energy of the electromagnetic field where $\hat{\mathbf{f}}^{\dagger}$ ($\hat{\mathbf{f}}$) is the LSP polaritonic vector field operator associated to the creation (annihilation) of a quantum of electromagnetic mode in presence of the MNP.
The last term describes the emitter-field interaction under the rotating-wave approximation. 

The electromagnetic field has to be quantized by taking into account the dispersing and absorbing nature of the metal \cite{Knoll-Schell-Welsch:2001,vanVlack-Hughes:2014,Hakami-Wang-Zubairy:2014}. 
The electromagnetic mode dispersion and absorption are governed by the real and imaginary parts of the metal dielectric constant 
$\varepsilon_m(\mathbf{r},\omega)=\varepsilon_R(\mathbf{r},\omega)+i\varepsilon_I(\mathbf{r},\omega)$, that satisfy the Kramers-Kronig relations. As a situation, we assume a Drude-like behaviour 
$\varepsilon_m(\omega)=\varepsilon_{\infty}-\omega_p^2/(\omega^2+i\gamma_p\omega)$\cite{KKnote}. 
We use $\varepsilon_{\infty}=6$, $\hbar\omega_p=7.90$ eV and $\hbar\gamma_p=51$ meV for silver \cite{vanVlack-Hughes:2014}.  

The quantization can be performed by introducing a noise polarization operator expressed in terms of the creation (annihilation) operators $\mathbf{\hat{f}}^{\dagger}$ ($\mathbf{\hat{f}}$) \cite{Knoll-Schell-Welsch:2001}. The electric field operator can be expressed as $\mathbf{\hat{E}}(\mathbf{r})=\mathbf{\hat{E}}^{(+)}(\mathbf{r})+\mathbf{\hat{E}}^{(-)}(\mathbf{r})$ with 
\begin{align}
\mathbf{\hat{E}}^{(+)}(\mathbf{r})=&\int_0^{\infty}d\omega\ \mathbf{\hat{E}}(\mathbf{r},\omega)  {,} \hspace{0.3cm} \mathbf{\hat {E}}^{(-)}(\mathbf{r})=[\mathbf{\hat{E}}^{(+)}(\mathbf{r})]^{\dag},\nonumber\\
\mathbf{\hat{E}}(\mathbf{r},\omega)=& i\sqrt{\frac{\hbar}{\pi\epsilon_0}}k_0^2
\int d{\mathbf{r}'} \sqrt{\varepsilon_I(\mathbf{r}',\omega)}
{\mathbf G}({\mathbf r},{\mathbf r}',\omega)\hat{{\mathbf f}}({\mathbf r'},\omega),
\end{align}
where $k_0=\omega/c$ and $\mathbf{G}$ is the Green's tensor. It contains all the information about the field response of the MNP.  
%

In the following, we investigate the optical response of the emitter-MNP system. We assume an emitter initially in its excited state $\ket{e}$ and the LSP field in the ground state (vacuum). The wave function of the hybrid system in the interaction picture writes at time $t$ \cite{vanVlack-Hughes:2014,Hakami-Wang-Zubairy:2014}
\begin{eqnarray}
&&\ket{\psi(t)} = \ C_e (t)e^{-i\omega_{e} t}\ket{e}\ket{\varnothing}\\
\nonumber
&&+ \int d\mathbf{r}\int_0^{\infty}d \omega\ e^{-i(\omega+\omega_g) t}\mathbf{C}_g(\mathbf{r},\omega,t)\cdot\ket{g}\ket{\mathbf{1}(\mathbf{r},\omega)} 
\label{wavefun}
\end{eqnarray}
where $\ket{e}\ket{\varnothing}$ corresponds to the emitter in its excited state and no LSP mode excited whereas
$\ket{g}\ket{\mathbf{1}(\mathbf{r},\omega)}$ corresponds to the emitter in its ground state and a single excited LSP mode of energy $\hbar \omega$. The elementary excitation of a LSP  is defined through the action of the bosonic vector field operator on the vacuum state $\mathbf{f}^{\dag}(\mathbf{r},\omega)\ket{\varnothing}=\ket{\mathbf{1}(\mathbf{r},\omega)}$. The dynamics of the probability amplitudes $C_{e}(t)$ and $\mathbf{C}_{g}(t)$ are derived from the time-dependent Schr{\"o}dinger equation \cite{vanVlack-Hughes:2014,Hakami-Wang-Zubairy:2014}.

As a first step, the coupling between the emitter and the MNP can be studied in the near field through the polarization spectrum $P(\omega)=\langle\hat{\sigma}^{\dagger}_{ge}(\omega)\hat{\sigma}_{ge}(\omega)\rangle$ \cite{vanVlack-Hughes:2014}
\begin{eqnarray}
&&P(\omega)=
\left|\frac{1}{\omega_{eg}-\omega-i\frac{\gamma_{d}}{2}-\frac{k_0^2}{\hbar\epsilon_0}d_{eg}^2 G_{uu}^{scatt}(\mathbf{r}_d,\mathbf{r}_d,\omega)}\right|^2
\label{polarisation_finale}
\end{eqnarray} 
 $G_{uu}$ is the dyadic component along  the direction $\mathbf{u}$ of the dipolar emitter ($\mathbf{d}_{eg}=d_{eg}\mathbf{u}$). Note that the free-space contribution of the Green's tensor is included in the transition frequency $\omega_{eg}$ (Lamb shift) and decay rate $\gamma_d$ (Weisskopf-Wigner theory). Therefore, only the scattering part of the Green's tensor appears in Eq. (\ref{polarisation_finale}).

The polarization spectrum characterizes the near field emission properties of the coupled system. It is also necessary to define the signal radiated in the far field zone.  Following Ref. \cite{Hakami-Wang-Zubairy:2014}, the  spectrum recorded at the detector position $\mathbf{r}$ expresses
\begin{eqnarray}
&&S(\mathbf{r},\omega)=\nonumber\\
\nonumber
&&\frac{1}{2\pi}\int_0^{\infty}\,dt_2 \int_0^{\infty}\,dt_1\,e^{-i\omega(t_2-t_1)}\langle\mathbf{\hat{E}}^{(-)}(\mathbf{r},t_2)\cdot\mathbf{\hat{E}}^{(+)}(\mathbf{r},t_1)\rangle \\
&&=
\frac{1}{2\pi}\left|\frac{k_0^2}{\hbar\epsilon_0} d_{eg}^2 G_{uu}(\mathbf{r},\mathbf{r}_d,\omega)\right|^2 P(\omega).
\end{eqnarray}

\begin{figure}[h]
	\includegraphics[width=0.47\textwidth]{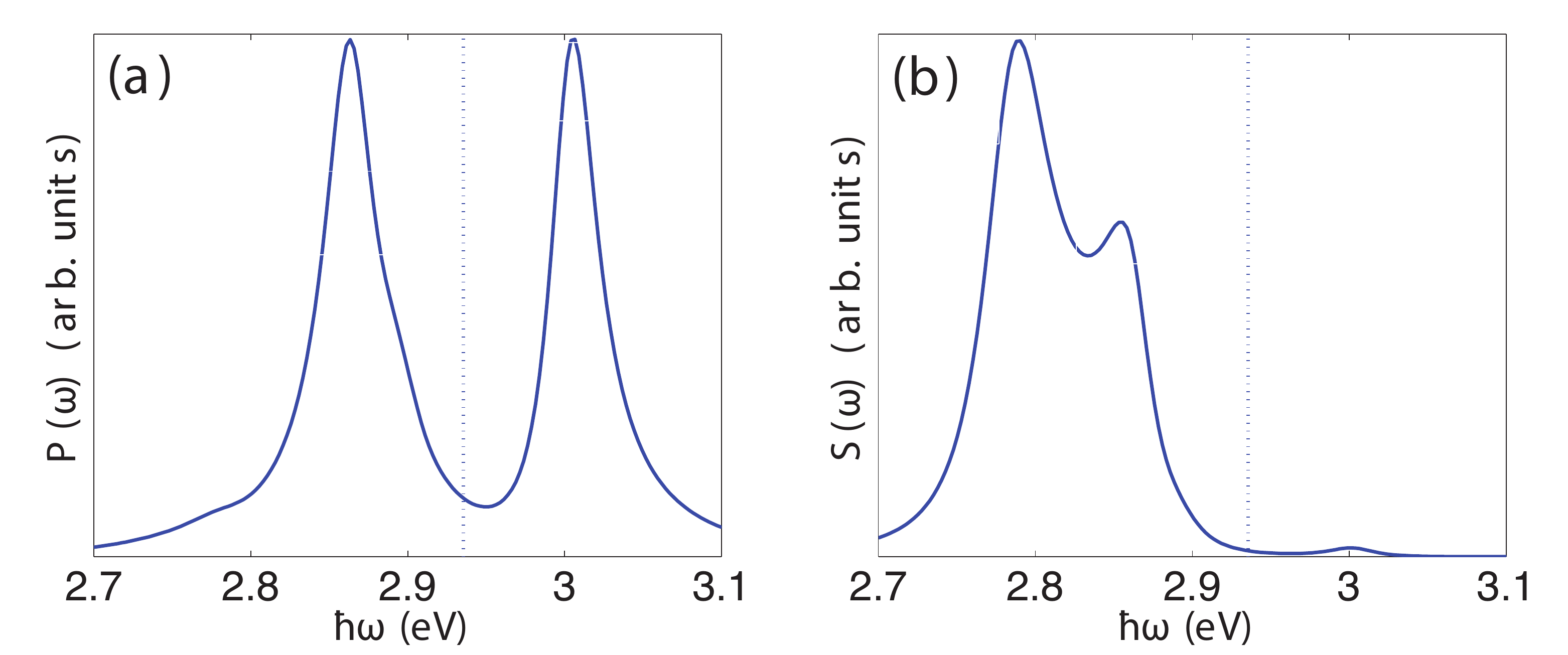}\\
		\includegraphics[width=0.47\textwidth]{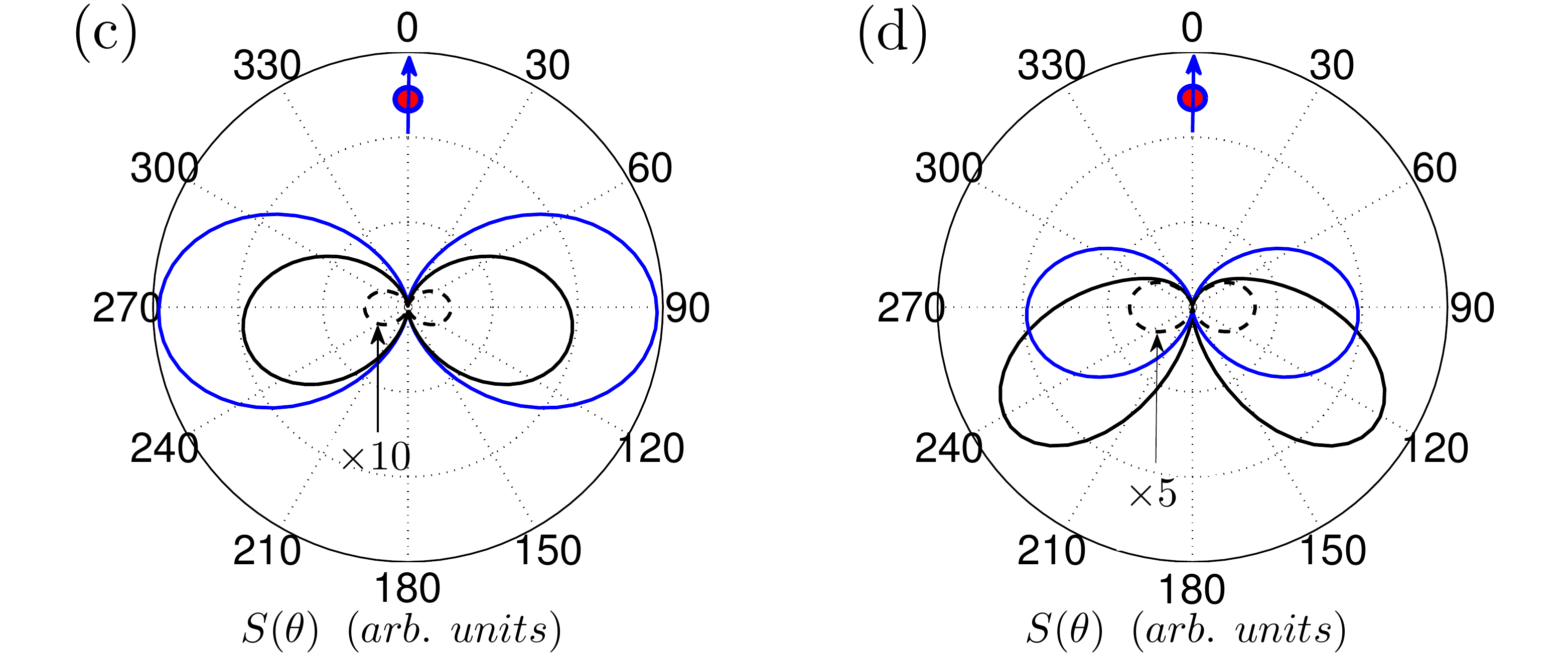}
	\caption{Top line: Near (a) and far (b) field spectra calculated for an emitter located 2 nm away from a MNP of radius $R=8$ nm, including all the MNP modes (N=25 ensures the convergence). The detector is located at (r=$1 \mu m$,$\theta=\pi/2$). The vertical line refers to the emitter emission frequency ($\hbar\omega_{eg}=2.94$ eV). Bottom line: Recorded signal as a function of the detector position for a sphere radius $R=8$ nm (c) and $R=20$nm (d). The three curves are calculated at the three peaks of the far field spectrum $S(\omega)$. Blue curves: $\hbar\omega=2.79$ eV (c) or $2.76$ eV (d). Black solid curves: $\hbar\omega=2.86$ eV (a) or $2.89$ eV (b). Black dashed curves: $\hbar\omega=3$ eV (c) or $3.02$ eV (d).
\label{spectre}}
\end{figure}

Both expressions can be applied in the weak and strong coupling limits since no Markov approximation has been made. For comparison purpose, we consider an emitter with $\hbar\gamma_d=15$ meV and dipole moment$\vert \mathbf{d}_{eg}\vert=24$ D, radially oriented, as in Ref. \cite{vanVlack-Hughes:2014,Hakami-Wang-Zubairy:2014}, see Fig. \ref{spectre}a. The polarization (near field) spectrum $P(\omega)$ presents a split of $\hbar\Delta\omega=144$ meV, revealing the strong coupling regime. The signal recorded in the far field is radically different, see Fig. \ref{spectre}b). We observe three peaks: the dominant one at 2.8 eV is associated to the LSP dipolar resonance (denoted LSP$_1$), known to be strongly radiative. The two others peaks show reminiscence of the mode splitting, with one of them dominant (near $2.85$ eV) and the last one hardly observable (near 3 eV). These 3 peaks can present similar amplitudes for large particles. We also represent the radiation diagram in Fig. \ref{spectre}c,d for two particle radii. We recover the dipolar angular emission for all the wavelength [$S(\theta)\propto \sin \theta$] except at the emission energy close to $\hbar\omega=2.9$ eV for which forward scattering occurs for the largest particle (solid black curve, Fig. \ref{spectre}d). This reveals the role of the quadrupolar mode (LSP$_2$) in the coupling process \cite{Bohren-Huffman:1983}.

More understanding of the emitter-MNP coupling process is achievable using the effective model we recently developed \cite{Rousseaux-GCF:2016,Dzsotjan-GCF:2016}. In particular, it makes a complete analogy with cQED description, paving the way towards direct transposition of cavity controlled dynamics at the nanoscale. To this aim, the hybrid emitter-MNP system is described as a quantum emitter coupled to a reservoir of $N$ LSP modes structured by the local density of states. The interaction Hamiltonian can be written as 
\begin{eqnarray}
&&\hat H_I=i\hbar\int_0^{+\infty}\!\!\!\!\!\!\! d\omega \sum_{n=1}^N\kappa_n^*(\omega,\mathbf{r}_d)\hat{b}_{\omega,n}(\mathbf{r}_d)^{\dagger}\hat{\sigma}_{ge}-h.c. \;, \label{hamil2}\\
&&|\kappa_n(\omega,\mathbf{r}_d)|^2=\frac{k_0^2}{\hbar\pi\epsilon_0}\mathbf{d}_{eg}\cdot Im[\mathbf{G}_n(\mathbf{r}_d,\mathbf{r}_d,\omega)]\mathbf{d}_{eg}^\star \;\label{lien}
\end{eqnarray}
$\kappa_n$ quantifies the coupling between the emitter and the MNP $n^{th}$ mode. It is expressed in terms of the Green's dyad, linking the preceding description with the following effective model. The excitation of a single LSP  of order $n$ (LSP$_n$) obeys $\ket{1_{\omega,n}}=\hat{b}_{\omega,n}^{\dagger}(\mathbf{r}_d)\ket{\varnothing}$ with the bosonic operator $\hat{b}_{\omega,n}(\mathbf{r}_d)^{\dagger}=\mathbf{d}_{eg}\cdot \mathbf{\hat E}_{n}(\mathbf{r}_d,\omega)/\hbar\kappa_n$, $\mathbf{\hat E}_{n}$ is the electric field associated to mode $n$.
%
%

\begin{figure}[h!]
 \hspace{-0.15cm}\includegraphics[width=0.25\textwidth]{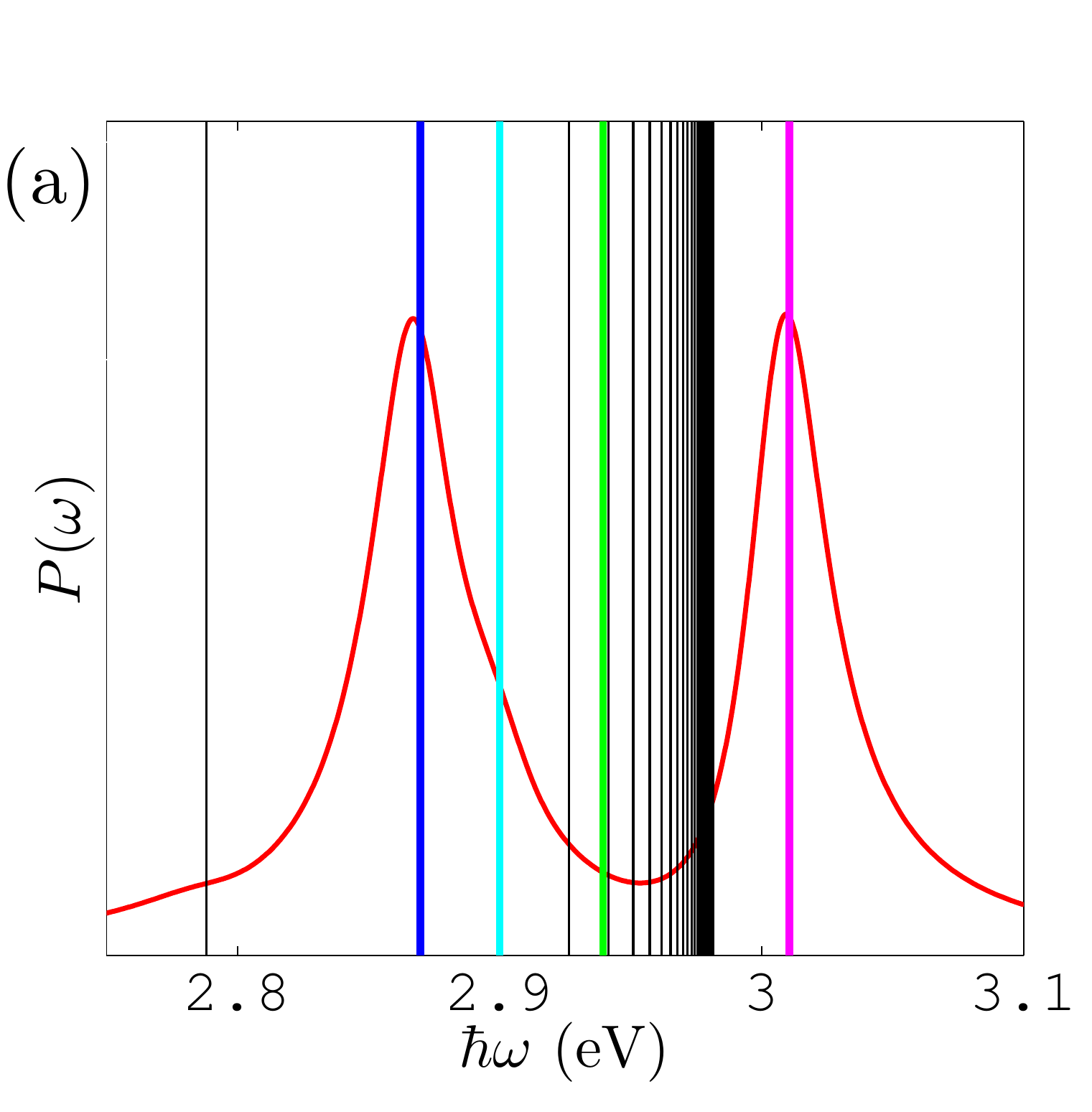}
  \hspace{-0.17cm}\includegraphics[width=0.28\textwidth]{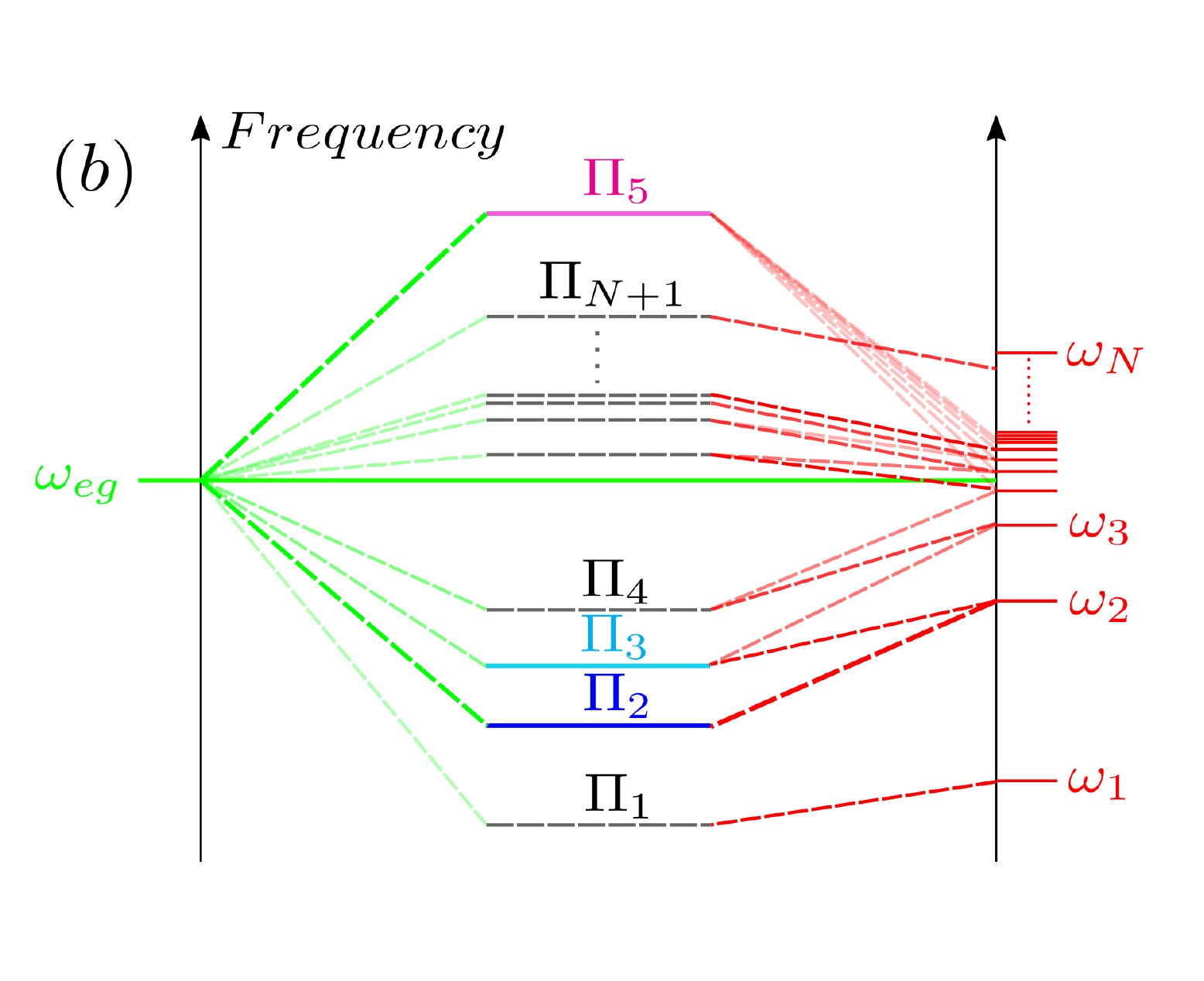}
	\caption{(a) Polarization spectrum (Fig. \ref{spectre}a). Black lines indicate the 26 hybrid modes. (b) Energy diagram of the hybrid system deduced from the full effective Hamiltonian diagonalization. A thicker line indicates a stronger weight $m_0$ of the atom  ($\vert e\rangle\vert\varnothing\rangle$, left part) or $m_n$ of the LSP$_n$ mode ($\vert g\rangle\vert 1_n\rangle$, right part of the diagram). In (a) the green line corresponds to the emission frequency of the emitter ($\hbar\omega_{eg}=2.94$ eV) leading to the strong coupling. The blue (magenta) line refers to the frequency $\Omega_2$ ($\Omega_5$) of the dressed state $\Pi_2$ ($\Pi_5$). The cyan line near $\hbar\omega\approx 2.9$ eV indicates the frequency of the $\Pi_3$ dressed state.\label{degen_multi}}
\end{figure}

Each resonance follows a lorentzian profile so that the coupling constant with a given mode can be represented by 
\begin{align}
\kappa_n(\omega,\mathbf{r}_d)=\sqrt{\frac{\gamma_n}{2\pi}}\frac{g_n(\mathbf{r}_d)}{\omega-\omega_n+i\frac{\gamma_n}{2}} \,,\label{structuration}
\end{align}
where $g_n$ is the coupling strength of the emitter to the MNP $n^{th}$ mode. $\omega_n$ and $\gamma_n$ are the mode resonance frequency and width, respectively. $\omega_n$ and $\gamma_n$ depend on the MNP material and size whereas the coupling strength $g_n$ depends also on the distance to the surface. We calculated that the coupling strength to a given mode fastly decay with distance, but can overcome the Joule losses in the MNP ($g_n >\gamma_p$) for separation distances below few nanometers, suggesting the feasibility of strong coupling. Specifically, we observe that high order modes play a significant role.
Finally, the effective Hamiltonian is obtained by tracing out the continuous degrees of freedom of the modes in order to establish a set of $N$ discrete modes. In the tensor product basis $\{\vert e\rangle\vert\vac\rangle,\vert g\rangle\vert 1_1\rangle,\cdots,\vert g\rangle\vert 1_N\rangle \}$, its matrix representation is \cite{Rousseaux-GCF:2016}
\begin{align}
H_{eff}=\hbar\begin{bmatrix}-i\frac{\gamma_{d}}{2} & ig_1 & ig_2 & \cdots & ig_N\\
-ig_1 & \Delta_1-i\frac{\gamma_1}{2} & 0 & \cdots & 0\\
-ig_2 & 0 & \Delta_2-i\frac{\gamma_2}{2} & \ddots & \vdots\\
\vdots & \vdots & \ddots & \ddots & 0\\
-ig_N & 0 & \cdots & 0 & \Delta_N-i\frac{\gamma_N}{2}\end{bmatrix},\label{eff_hamil}
\end{align}
where $\Delta_n=\omega_n-\omega_{eg}$ is the detuning from the resonance. This effective Hamiltonian provides a very practical representation of the hybrid configuration. The emitter couples to each LSP$_n$ mode with the coupling strength $g_n$. The losses $\gamma_n$ reflect the population leakage from the excited state $\vert g\rangle\vert 1_n\rangle$ to the ground state $\vert g\rangle\vert \varnothing\rangle$. In order to interpret the degeneracy breaking in the strong coupling regime (Fig.\ref{spectre}a,c), let us first consider the interaction between the quantum emitter and one single LSP mode of the MNP. Detailed analysis reveals that the third mode (LSP$_3$) presents the main contribution to the coupling process. Therefore, we approximate the effective Hamiltonian by
\begin{align}
H_{eff}\approx\hbar\begin{bmatrix}-i\frac{\gamma_{d}}{2} & ig_3\\
-ig_3 & \Delta_3-i\frac{\gamma_3}{2}\\
\end{bmatrix} \,.\label{eff_hamil_third}
\end{align}
If we neglect the loss rates $\gamma_{d}$ and $\gamma_3$, the diagonalization of the effective Hamiltonian leads to the dressed state of the hybrid emitter-LSP$_3$ system with angular frequencies 
$\Omega_{\pm}=(\omega_{eg}+\omega_{3})/2\pm\sqrt{g_3^2+\Delta_3^2/4} 
$.
\begin{figure}
\includegraphics[width=0.23\textwidth]{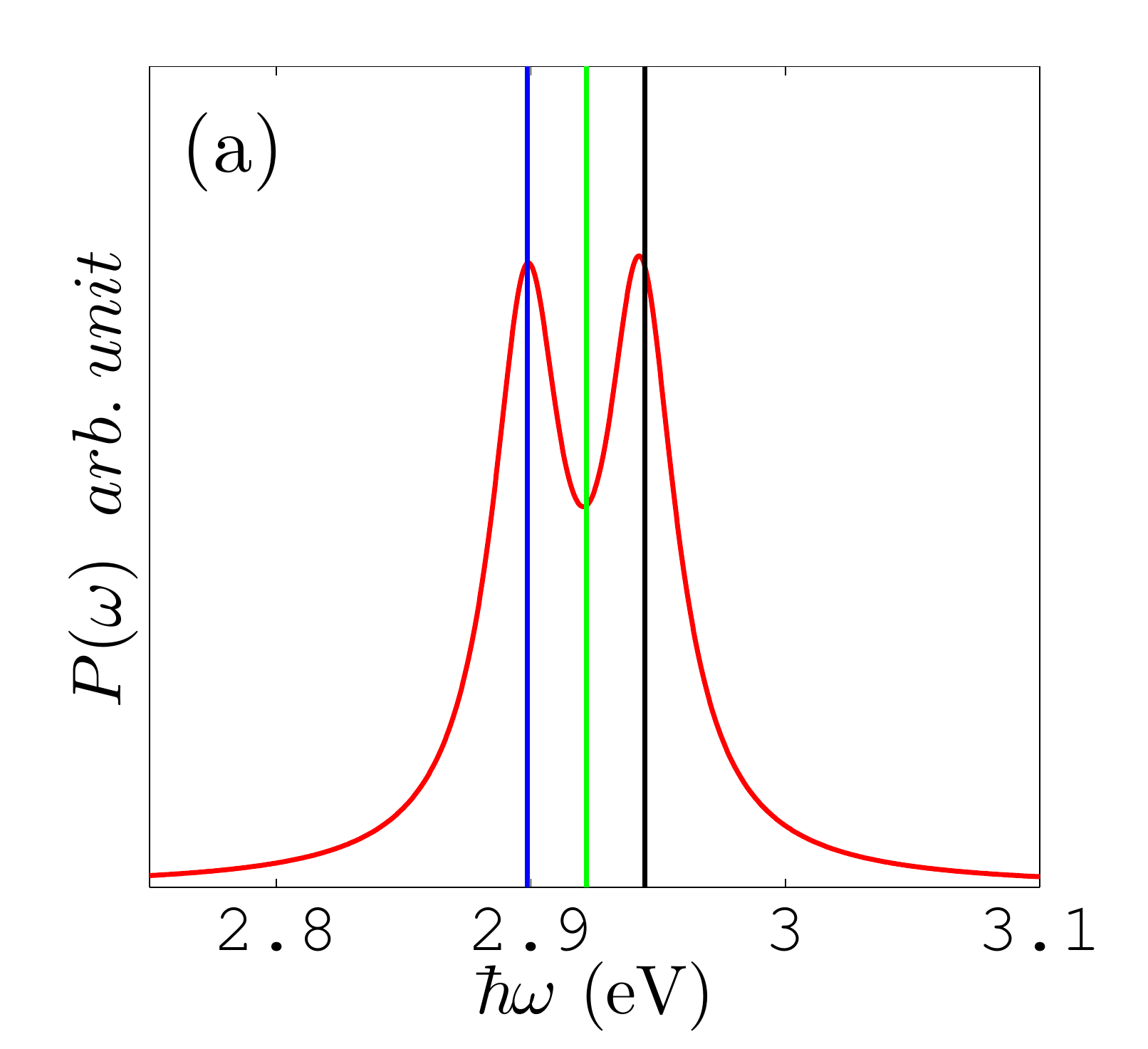}
\raisebox{0.78cm}{\includegraphics[width=0.23\textwidth]{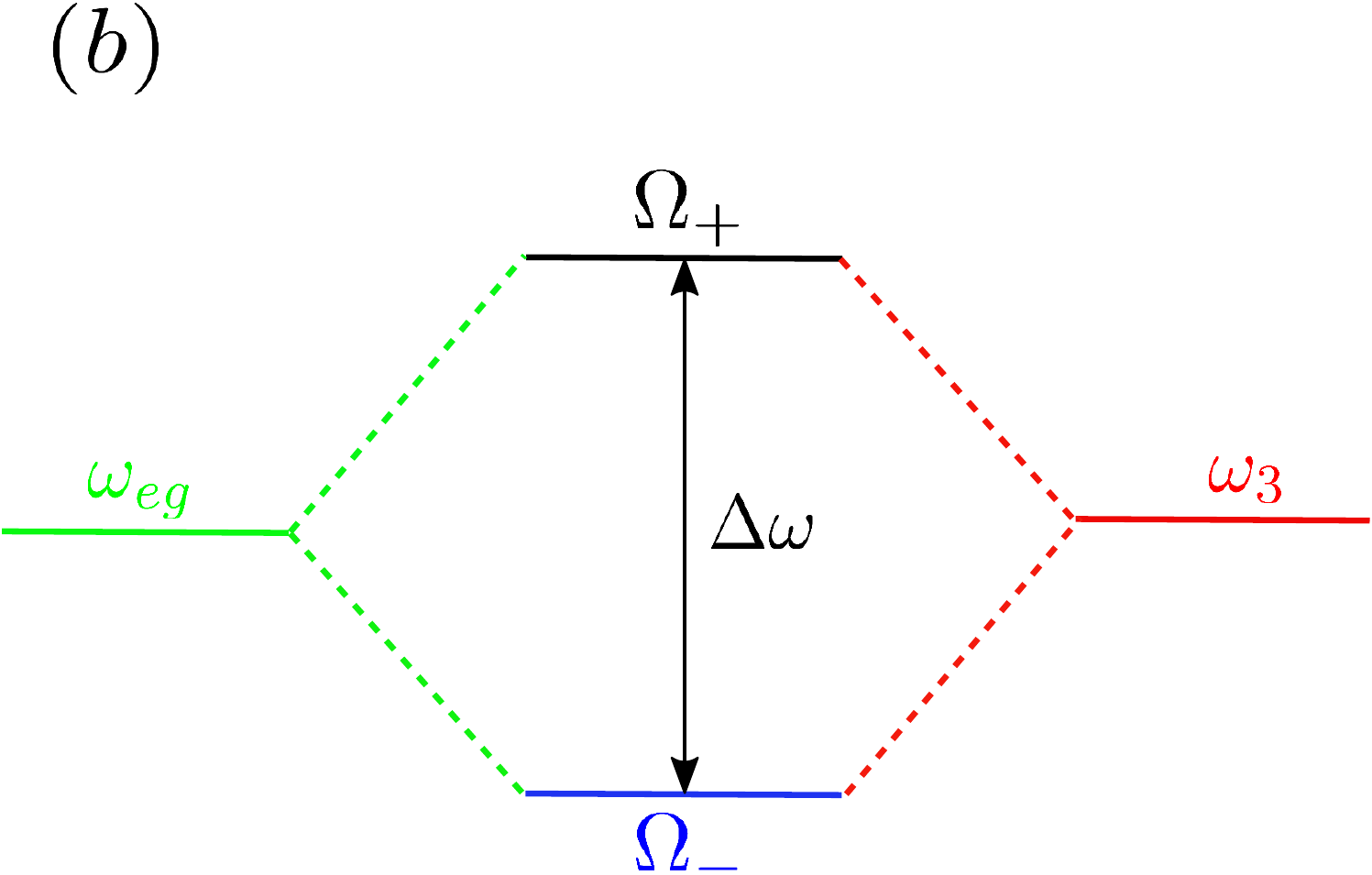}}
\caption{(a) Polarization spectrum keeping the LSP$_3$ mode contribution only. The emitter is resonant with the third mode ($\hbar\omega_{eg}=\hbar\omega_3=2.92$ eV, green line). The blue and black lines characterize the frequencies $\Omega_+$ and $\Omega_-$ of the two dressed states calculated from the approximated effective Hamiltonian \ref{eff_hamil_third}. (b) Energy diagram of hybrid nanosource.\label{degen_3}}
\end{figure}
If the emission is resonant with the LSP$_3$ mode ($\omega_{eg}=\omega_3$ so that $\Delta_3=0$), the energy splitting is $\hbar\Delta\omega=\hbar(\Omega_+-\Omega_-)=2\hbar g_3=47$ meV, that is close to the splitting observed in  the polarization spectrum calculated in Fig.\ref{degen_3}a)  ($\hbar\Delta\omega=43$ meV), where only the LSP$_3$ mode is considered. Taking into account the dissipation of the modes, the angular frequencies of the dressed states become $\Omega_\pm=\omega_{eg}+\mathfrak{Re}[\lambda_\pm]$
where $\lambda_\pm$ are the complex eigenvalues of the effective Hamiltonian (Eq. \ref{eff_hamil_third}). We recover $\hbar\Delta\omega=43$ meV, as expected. However, the energy splitting is still low compared to the one observed in Fig. \ref{spectre}a) ($\hbar\Delta\omega=144$ meV), that takes into account all the LSP modes of the MNP . We therefore consider all the 25 LSP modes in the effective Hamiltonian (Eq. \ref{eff_hamil}). Its diagonalisation leads to $26$ dressed states with angular frequencies $\Omega_m=\omega_{eg}+\mathfrak{Re}[\lambda_m] \,, (m=1,\ldots , 26)$
($\lambda_m$ are the eigenvalues). 
%
For such dissipative systems, we have to define right and left eigenvectors $\vert\Pi^R_m\rangle$ and $\vert\Pi^L_m\rangle$, respectively, satisfying $H_{eff}\vert\Pi^R_m\rangle=\lambda_m \vert\Pi^R_m\rangle$ and $H_{eff}^\dagger\vert\Pi^L_m\rangle=\lambda_m^\star \vert\Pi^L_m\rangle$, $\langle\Pi^L_m\vert\Pi^R_m\rangle=\delta_{mn}$. For Hamiltonian of the form (\ref{eff_hamil}), one can simply connect them as follows \cite{Guerin:2010}
\begin{align}
\vert\Pi_m^R\rangle&=m_0\vert e\rangle\vert\varnothing\rangle+\sum_{n=1}^N m_n \vert g\rangle\vert 1_n\rangle, \\
\vert\Pi_m^L\rangle&=-m_0^\star\vert e\rangle\vert\varnothing\rangle+\sum_{n=1}^N m_n^\star \vert g\rangle\vert 1_n\rangle,
\end{align}
where $m_0$ and $m_n$ gives the weight of each mode $\vert e\rangle \vert \varnothing \rangle$ or $\vert g\rangle \vert 1_n\rangle$. 
We have now all the ingredients to interpret the polarization spectrum in the strong coupling regime (see Fig. \ref{degen_multi}). The mode hybridization, deduced from the diagonalization of the Hamiltonian is depicted in Fig. \ref{degen_multi}b). We indicate the main LSP modes involved for each dressed state. We observe that the energy of the dressed states $\Pi_2$ and $\Pi_5$ exactly match the two peaks in the polarization spectrum. These dressed states mainly result from the hybridization of the excited level of the emitter with either the LSP$_2$ and LSP$_3$ ($\Pi_2$) or LSP$_6$ to LSP$_{11}$ ($\Pi_5$) modes of the MNP. In addition, the shoulder visible in the polarization spectrum near $\hbar\omega\approx2.9$ eV originates from the dressed states $\Pi_3$. Note that the $\Pi_1$ and $\Pi_2$ states present a large contribution of the TLS or radiative LSP$_{1,2}$ modes that radiate in the far field zone (see Fig. \ref{spectre}b and \ref{spectre}c,d). On the contrary, the dressed state $\Pi_5$ appears as a dark mode in agreement with the far field spectrum calculated in Fig. \ref{spectre}b).

It is worthwile to note that the strong coupling regime 
can be achieved at the single molecule level thanks to cumulative effect of coupling to several $LSP$ modes. Another possibility would be to increase the number $N$ of emitters coupled to the MNP. Indeed, the effective Hamiltonian presents a similar structure than atom in a cavity so that we also expect a Rabi splitting proportional to $\sqrt{N}$.  Taking benefit of both the number of involved LSP modes in the coupling process and increasing the number of molecules would relax the strong coupling conditions.

\begin{figure}
\hspace{1cm}\includegraphics[width=0.3\textwidth]{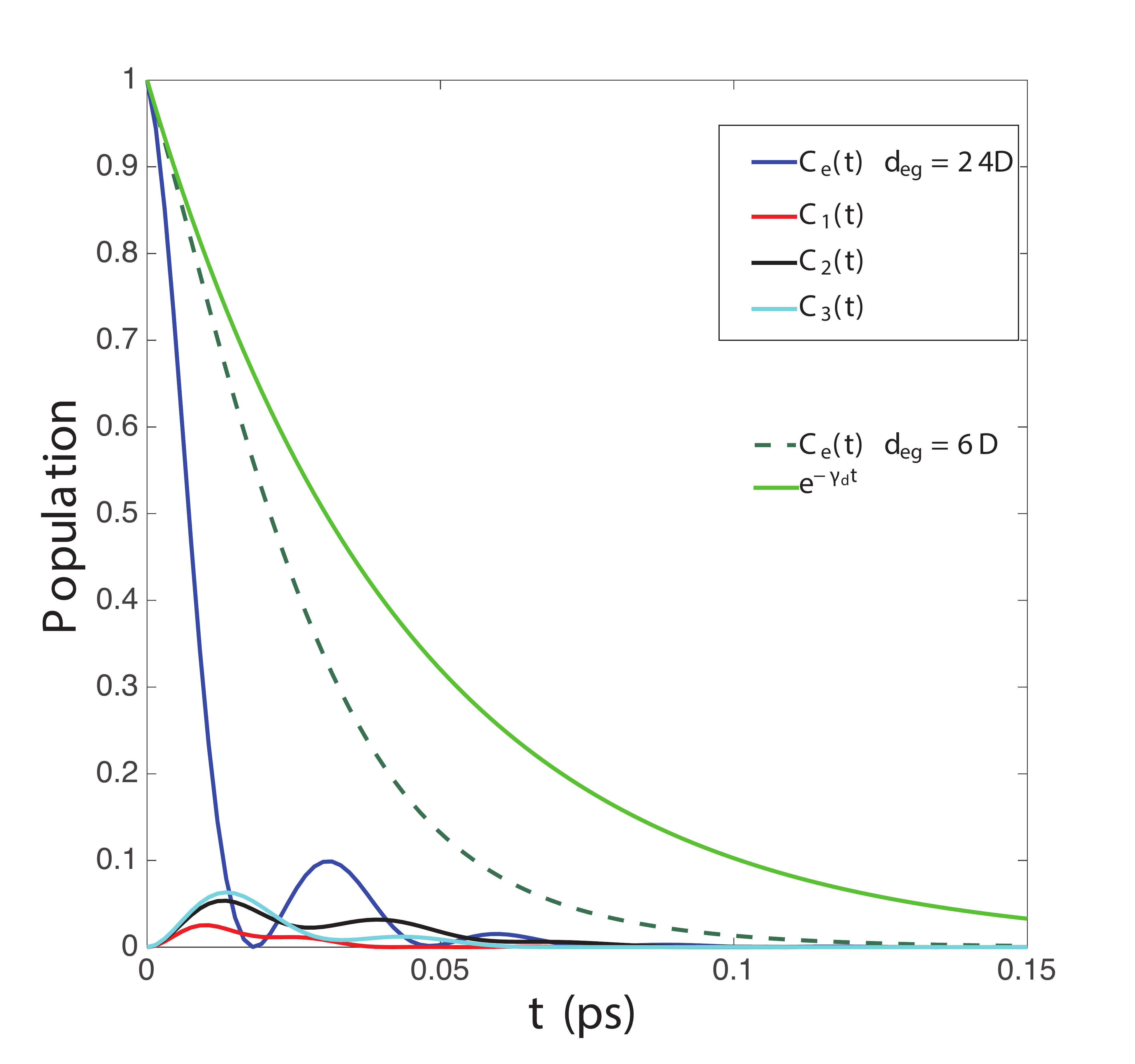}
\caption{Emitter and LSP population dynamics for different dipole moment value: $d=24$ D (strong coupling), $d=6$ D (showing a quasi-exponential decay) and in vacuum. \label{Dynamics}}
\end{figure}
The effective model also unravels the dynamics of the strongly coupled system. Indeed, the wavefunction writes at time $t$:  
$\ket{\psi(t)} = \sum_{m=1}^{N+1} \eta_m\vert \Pi_m^R \rangle e^{-i\lambda_m t} \,,$
with  $\eta_m=\langle\Pi_m^L \vert \psi(0)\rangle=-m_0$ if we assume an emitter initially in its excited state and no LSP mode populated. The evolution of the populations obey $\vert C_e(t)\vert^2=\vert\langle e,\varnothing  \vert\psi(t)\rangle\vert^2=\vert\sum_{m=1}^{N+1} m_0^2 e^{-i\lambda_m t}\vert^2$ for the excited state of the emitter and $\vert C_n(t) \vert^2=\vert\langle g,1_n  \vert \psi(t)\rangle\vert^2=\vert\sum_{m=1}^{N+1} m_0 m_n e^{-i\lambda_m t}\vert^2$  for the $n^{th}$ LSP mode. Figure \ref{Dynamics} presents the populations dynamics. Although strongly damped, a clear Rabi oscillation is visible revealing reversible ultrafast energy transfert with a period $T_{Rabi}=2\pi/\Delta \omega=0.03$ ps, as expected. The energy transfer between the emitter and the MNP is mainly governed by the LSP$_2$ and LSP$_3$ modes, the other being poorly populated.  Optimized configurations, such as nanoprism that facilitates the strong coupling regime \cite{Zengi-Kall-Shegai:2015} would permit to improve the energy transfer efficiency. Additionnally, we check that we recover a fast exponential decay in the weak coupling regime (for $d=6$ D), in agreement with the Fermi's golden rule.

To summarize, we have described the optical response of the hybrid MNP-quantum emitter nanosource in analogy with a cQED description and dressed atom picture. Specifically, we clarified the nature of the dressed states in the strong coupling regime. Since the effective Hamiltonian parameters are easily extracted from the Green's tensor of the plasmonic nanostructures, this formalism can be generalized to more complex system as {\it e.g.} plasmonic nanostructures of arbitrary shape, three-level system in $\Lambda$ configuration or adding an external driving field. Additionnally, this description offers a simple and very intuitive understanding of the spectroscopic properties of the hybrid nanosource. Finally, although relying on a different paradigm (mode confinement instead of mode lifetime), this formalism permits a direct transposition of cQED concept to the nanoscale and constitutes therefore a powerful tool to propose and design original nanophotonics or plasmonics devices. 
\section{acknowledgements}
We acknowledge support from the French National Agency: Labex ACTION (ANR-11-LABX-01-01) and PLACORE (ANR-BS10-0007), and from the Conseil
Regional de Bourgogne and FEDER (PARI-PHOTCOM). DD and GCF thank the european COST action MP1403 Nanoscale Quantum Optics.

\end{document}